\documentclass[twocolumn,pra,showpacs,floatfix]{revtex4}
\usepackage{amsmath}
\usepackage{graphicx}

\newcommand{\sinc}{\text{sinc}}
\newcommand{\arcsinc}{\text{arcsinc}}

\begin{document}
\title{Tackling Systematic Errors in Quantum Logic Gates with Composite Rotations}
\author{Holly K. Cummins}
\affiliation{Centre for Quantum Computation, Clarendon Laboratory,
University of Oxford, Parks Road, OX1 3PU, United Kingdom}
\author{Gavin Llewellyn}
\affiliation{Centre for Quantum Computation, Clarendon Laboratory,
University of Oxford, Parks Road, OX1 3PU, United Kingdom}
\author{Jonathan A. Jones}\email{jonathan.jones@qubit.org}
\affiliation{Centre for Quantum Computation, Clarendon Laboratory,
University of Oxford, Parks Road, OX1 3PU, United Kingdom}
%\affiliation{Oxford Centre for Molecular Sciences, Central
%Chemistry Laboratory, University of Oxford, South Parks Road, OX1
%3QH, United Kingdom}
\date{\today}
\begin{abstract}
We describe the use of composite rotations to combat systematic
errors in single qubit quantum logic gates and discuss three
families of composite rotations which can be used to correct
off-resonance and pulse length errors. Although developed and
described within the context of NMR quantum computing these
sequences should be applicable to any implementation of quantum
computation.
\end{abstract}
\maketitle

\section{Introduction}
Quantum computers \cite{bennett00} are information processing
devices that use quantum mechanical effects to implement
algorithms which are not accessible to classical computers, and
thus to tackle otherwise intractable problems \cite{shor99}.
Quantum computers are extremely vulnerable to the effects of
errors, and considerable effort has been expended on alleviating
the effects of random errors arising from decoherence processes
\cite{shor95, steane96, steane99}.  It is, however, also important
to consider the effects of systematic errors, which arise from
reproducible imperfections in the apparatus used to implement
quantum computations.

The effects of systematic errors are clearly visible in nuclear
magnetic resonance (NMR) experiments \cite{ernst87} which have
been used to implement small quantum computers \cite{cory96,
cory97, gershenfeld97, jones98c, jones01a, vandersypen01}.
Implementing complex quantum algorithms require a network of many
quantum logic gates, which for an NMR implementation translates
into even longer cascades of pulses. In these cases small
systematic errors in the pulses (which can be ignored in many
conventional NMR experiments) accumulate and have significant
effects.

It makes sense to consider systematic errors as some of them can
be tackled relatively easily. In the Bloch picture, where unitary
operations are visualized as rotations of the Bloch vector on the
unit sphere, systematic errors are expressed as rotational
imperfections. The sensitivity of the final state to these
imperfections can be much reduced by replacing single rotations
with composed rotations as discussed below.

\section{Systematic errors in NMR quantum computers}
Any implementation of a quantum computer requires quantum bits
(qubits) on which the quantum information is stored, and quantum
logic gates which act on the qubits to process the quantum
information.  Fortunately it is only necessary to implement a
small set of quantum logic gates, as more complex operations can
be achieved by joining these gates together to form logic
circuits.  A simple and convenient set comprises a range of single
qubit gates together with one or more two qubit gates, which
implement conditional evolutions and thus logical operations
\cite{barenco95}.

NMR quantum computers are implemented \cite{jones01a} using the
two spin states of spin-1/2 atomic nuclei in a magnetic field as
the qubits. Transitions between these states, and thus single
qubit gates, are achieved by the application of radio frequency
(RF) pulses.  Two qubit gates require some sort of spin--spin
interaction, which in NMR is provided by the scalar spin--spin
coupling ($J$ coupling) interaction.  While this does not have
quite the form needed for standard two qubit gates, it can be
easily sculpted into the desired form by combining free evolution
under the background Hamiltonian (which includes spin--spin
coupling terms) with the application of single qubit gates
\cite{jones01a}.

As single qubit gates involve the application of external fields
they are vulnerable to systematic errors in these fields.  In the
ideal case, the application of a RF field in resonance with the
corresponding transition with relative phase $\phi$ (in the
rotating frame \cite{ernst87}) will drive the Bloch vector through
some angle about an axis orthogonal to the $z$-axis and at an
angle $\phi$ to the $x$-axis. The rotation angle, $\theta$,
depends on the nutation rate induced by the RF field, usually
written $\nu_1$, and the duration of the pulse, $\tau$. In
practice the RF field is not ideal, and this leads to two
important types of systematic errors, pulse length errors and
off-resonance effects \cite{ernst87,freeman97b}.

Pulse length errors occur when the duration of the RF pulse is set
incorrectly, or (equivalently) when the RF field strength deviates
from its nominal value, so that the rotation angle achieved
deviates from its theoretical value.  Within NMR this effect is
most commonly observed as a result of spatial inhomogeneity in the
applied RF field, so that it is impossible for all the spins
within a macroscopic sample to experience the same rotation angle.
Off-resonance effects arise when the RF field is not quite in
resonance with the relevant transition, so that the rotation
occurs around some tilted axis.

Composite pulses \cite{ernst87,freeman97b,levitt86} are widely
used in NMR to minimize the sensitivity of the system to these
errors by replacing simple rotations with composite rotations
which are less susceptible to such effects. However, conventional
composite pulse sequences are rarely appropriate for quantum
computation because they usually incorporate assumptions about the
initial state of the spins.  Such starting states are not known
for pulses in the middle of complex quantum computations, and it
is therefore necessary to use fully-compensating (type A)
composite pulse sequences \cite{levitt86}, which work for any
initial state. Composite pulses of this kind, which do not offer
quite the same degree of compensation as is found with more
conventional sequences, are of little use in conventional NMR, and
have received relatively little study.  They are, however, ideally
suited to quantum computation.

\section{Off-resonance Errors}
The problem of tackling off-resonance errors was initially studied
by Tycko \cite{tycko83}; his results were then extended by Cummins
and Jones \cite{cummins00, cummins01}.  Here we describe two
families of composite pulses which can be used to compensate for
off-resonance errors, and show how they can be derived using
quaternions.

The original method used to develop many type A composite pulse
sequences \cite{tycko83, cummins00, cummins01} was based on
dividing the propagator describing the evolution of the quantum
system into intended and error components, and then seeking to
minimise the error term.  While this approach is effective, it is
cumbersome, and a much simpler approach can be adopted for single
qubit gates, which are simply rotations on the Bloch sphere and so
can be modelled by quaternions.  The quaternions corresponding to
individual pulses can be multiplied together to give a quaternion
description of the composite pulse, which can then be compared
with the quaternion of the ideal system.

A quaternion is often thought of as a vector with four
coefficients, but when describing a rotation it is more useful to
regroup these coefficients as a scalar and a three-vector,
\begin{equation}
\mathsf{q}=\{s, \mathbf{v}\}
\end{equation}
where
\begin{equation}
s=\cos(\theta/2)
\end{equation}
depends solely on the rotation angle, $\theta$, and
\begin{equation}
\mathbf{v}=\sin(\theta/2)\mathbf{a}
\end{equation}
depends on both the rotation angle, $\theta$, and a unit vector
along the rotation axis, $\mathbf{a}$.  Thus the quaternion
describing an on-resonance pulse with phase angle $\phi$ is
\begin{equation}
\mathsf{q}_{\theta\phi}=\{\cos(\theta/2),
\sin(\theta/2)\{\cos(\phi),\sin(\phi),0\}\}.
\end{equation}
An off-resonance pulse is conveniently parameterised by its
off-resonance fraction $f=\delta/\nu_1$ (where $\delta$ is the
off-resonance frequency, and $\nu_1$ the nutation rate), and is
described by the quaternion
\begin{equation}
\mathsf{q}_{\theta\phi}=\{\cos(\theta'/2),
\frac{\sin(\theta'/2)}{\sqrt{1+f^2}}\{\cos(\phi),\sin(\phi),f\}\}
\end{equation}
where $\theta'=\theta\sqrt{1+f^2}$, and $\theta$ is now the
nominal rotation angle, that is the rotation achieved when $f=0$.
The quaternion describing a sequence of pulses is obtained by
multiplying the quaternions for each pulse according to the rule
\begin{equation}
\mathsf{q_1}\ast\mathsf{q_2}=\{s_1\cdot s_2
-\mathbf{v_1}\cdot\mathbf{v_2},
s_1\mathbf{v_2}+s_2\mathbf{v_1}+\mathbf{v_1}\wedge\mathbf{v_2}\}.
\end{equation}
Finally, two quaternions can be compared using the quaternion
fidelity \cite{levitt86}
\begin{equation}
\mathcal{F}(\mathsf{q_1},\mathsf{q_2})=|\mathsf{q_1}\cdot\mathsf{q_2}|
=|{s_1}\cdot{s_2}+\mathbf{v_1}\cdot\mathbf{v_2}|
\label{eq:fidelity}
\end{equation}
(it is necessary to take the absolute value, as the two
quaternions $\{s,\mathbf{v}\}$ and $\{-s,-\mathbf{v}\}$ correspond
to equivalent rotations, differing in their rotation angle by
integer multiples of $2\pi$).

Following our previous work \cite{cummins00} we seek to tackle
off-resonance errors in a $\theta_x$ pulse using a sequence of
three pulses applied along the $x$, $-x$ and $x$ axes; pulses with
any other phase angle can then be trivially derived by simply
adding the desired value to the phase angles of all the pulses in
the sequence. Such sequences can be described completely by the
nominal rotation angles of the three pulses, $\theta_1$,
$\theta_2$ and $\theta_3$. The composite quaternion for this
composite pulse is complicated, but the situation can be greatly
simplified by expanding it as a Maclaurin series in $f$ and
neglecting all terms above the first power. This gives
\begin{equation}
s=\cos\left(\frac{\theta_1-\theta_2+\theta_3}{2}\right)
\end{equation}
and
\begin{multline}
\mathbf{v}=\left\{\sin\left(\frac{\theta_1-\theta_2+\theta_3}{2}\right)\right.,\;
\sin\frac{\theta_2}{2}\sin\left(\frac{\theta_1-\theta_3}{2}\right),\\
\left.f\left(2\cos\left(\frac{\theta_2-\theta_3}{2}\right)\sin\frac{\theta_1}{2}-
\sin\left(\frac{\theta_1-\theta_2-\theta_3}{2}\right)\right)\right\},
\end{multline}
while the ideal quaternion has the form
\begin{equation}
\{\cos(\theta/2), \{\sin(\theta/2),0,0\}\}.
\end{equation}
It now remains to chose the three nominal rotation angles so that
these equations agree.

First we note that in order to achieve the correct rotation angle,
$s=\cos(\theta/2)$, we must choose our angles such that
$\theta_1-\theta_2+\theta_3=\theta+2a\pi$ (where $a$ is any
integer). We also note that the $y$ component of $\mathbf{v}$
should equal zero, and that this can be achieved by choosing
$\theta_1=\theta_3+2b\pi$ (where $b$ is any integer). These two
choices give
\begin{equation}
\mathbf{v}=\left\{\sin\frac{\theta}{2},\;0,\;
f\left(\sin\frac{\theta}{2}-2\sin\left(\frac{\theta}{2}-\theta_1\right)\right)\right\}.
\end{equation}
Finally we choose $\theta_1$ such that the $z$ component of
$\mathbf{v}$ equals zero; this gives
\begin{equation}
\theta_1=\frac{\theta}{2}-\arcsin\left(\frac{\sin(\theta/2)}{2}\right).
\end{equation}
Combining this value with our previous relations between the
angles gives
\begin{align}
\theta_1&=2n_1\pi+\frac{\theta}{2}-\arcsin\left(\frac{\sin(\theta/2)}{2}\right)\\
\theta_2&=2n_2\pi-2\arcsin\left(\frac{\sin(\theta/2)}{2}\right)\\
\theta_3&=2n_3\pi+\frac{\theta}{2}-\arcsin\left(\frac{\sin(\theta/2)}{2}\right)
\end{align}
where $n_1$, $n_2$ and $n_3$ are integers, subject to the physical
restriction that the resulting pulse angles must be positive.

These solutions have the same general form as those found
previously \cite{cummins00}.  Although they appear to differ in
detail the expressions are, in fact, identical: taking the values
$n_1=1$, $n_2=1$ and $n_3=0$ gives our previous family of
solutions \cite{cummins00}, referred to by the acronym
\textsc{corpse} (Compensation for Off-Resonance with a Pulse
SEquence).  This family is now seen to be just one member of a
larger group of families.  To choose between these it is necessary
to look at higher order terms, and this is most conveniently
achieved using the quaternion fidelity,
Equation~\ref{eq:fidelity}. As a baseline we take the fidelity of
a single off-resonance $\theta_\phi$ pulse compared with its
(ideal) on-resonance form,
\begin{equation}
\mathcal{F}\approx 1+f^2\left(\frac{\cos\theta-1}{4}\right)
\label{eq:fidfplain}
\end{equation}
where terms in $f^4$ and higher have been neglected.  Note that
the fidelity only contains even order terms in $f$ as the
composite pulse perform symmetrically for positive and negative
values of $f$.

As expected all members of our general group of solutions result
in much better fidelities; in particular the term in $f^2$ is
always completely removed.  The behaviour of the term in $f^4$ is
much more complicated, but it can be shown that this term depends
\emph{only} on the value of $n=n_1-n_2+n_3$, that is the total
number of \emph{additional} $2\pi$ rotations performed by the
composite pulse sequence, and has the smallest absolute value when
the three integers are chosen so that $n=0$. As our previous
values ($n_1=1$, $n_2=1$ and $n_3=0$) are the smallest numbers
that fit this criterion, it seems that the \textsc{corpse} family
of pulse sequences is indeed the best member of this group.  The
only other family of interest is that with $n_1=0$, $n_2=1$ and
$n_3=0$, previously referred to as \textsc{short-corpse}
\cite{cummins01}; while this performs less well than
\textsc{corpse} it is somewhat shorter.   Numerical values of
individual pulse rotation angles for \textsc{corpse} sequences
with a variety of target angles are given in
Table~\ref{tab:corpse}.
\begin{table}
\begin{tabular}{rrrr}\hline
$\theta$&$\theta_1$&$\theta_2$&$\theta_3$\\\hline
$30^\circ$&367.6&345.1&7.6\\
$45^\circ$&371.5&337.9&11.5\\
$90^\circ$&384.3&318.6&24.3\\
$180^\circ$&420.0&300.0&60.0\\\hline
\end{tabular}
\caption{Pulse rotation angles for a \textsc{corpse} composite
pulse with a target rotation of $\theta_x$; \textsc{corpse} pulse
phases are $+x$, $-x$, $+x$.}\label{tab:corpse}
\end{table}

\begin{figure}
\includegraphics[scale=0.2]{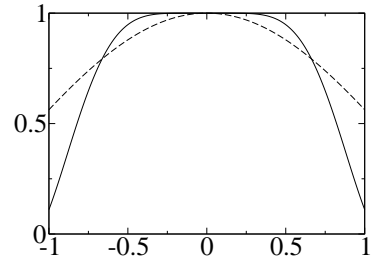}
\caption{Fidelity of simple (dashed line) and \textsc{corpse}
composite pulses (solid line) as a function of the off-resonance
fraction $f$ for pulses with a target rotation angle of
$180^\circ$.} \label{fig:corpse}
\end{figure}

The performance of the \textsc{corpse} sequence for a $180^\circ$
pulse is demonstrated in Figure~\ref{fig:corpse};
the \textsc{corpse} pulse performs better than a simple pulse as
long as $|f|\leq0.663$.
%\begin{equation}
%|f|\leq\sqrt{\left(\frac{6\arccos((1+\sqrt{5})/4)}{\pi}\right)-1}\approx0.663
%\end{equation}
For smaller values of $\theta$ the effective range of $f$ is
reduced, but not dramatically so: for a $30^\circ$ pulse the
\textsc{corpse} pulse outperforms a simple pulse as long as
$|f|\leq0.297$.

\section{Pulse length errors}
A similar approach can be used to develop composite pulses to
tackle pulse length errors.  As before we begin with a sequence of
three pulses, but the subsequent development is quite different.
In particular we allow the three pulses to have arbitrary phase
angles, as well as arbitrary rotation angles, thus giving us six
variable parameters, although this number is soon reduced to
three.

The quaternion corresponding to each pulse takes the simple form
\begin{equation}
\mathsf{q}_{\theta\phi}=\{\cos(\theta'/2),
\sin(\theta'/2)\{\cos(\phi),\sin(\phi),0\}\}.
\end{equation}
where $\theta'=\theta(1+g)$ is the \emph{actual} rotation angle
achieved by a pulse with nominal rotation angle $\theta$, and $g$
is the fractional error in the pulse power.  The quaternion
describing the composite pulse is very complicated, but can be
simplified by restricting attention to the time symmetric case,
where $\theta_1=\theta_3$ and $\phi_1=\phi_3$.  This automatically
ensures that the composite quaternion has no $z$ component, as any
time symmetric sequence of rotations about axes in the $xy$ plane
is itself a rotation about an axis in the $xy$ plane.

Even after this simplification, the composite quaternion remains
extremely complicated.  To make further progress we note that a
composite pulse of this kind has been previously described for the
case of a $180^\circ$ rotation: the sequence
\begin{equation}
180_{60}\,180_{300}\,180_{60} \label{eq:scrof180}
\end{equation}
will perform a $180_x$ rotation with compensation for pulse length
errors (see \cite{tycko84}, but note the corrected phase angles).
It seems likely that other members of this family will have
\emph{either} $\theta_1=\pi$ or $\theta_2=\pi$; both possibilities
were initially explored, but the second choice seemed more
productive and forms the basis of our subsequent work.

As before, the composite quaternion can be expanded as a Maclaurin
series in $g$, and it is most useful to concentrate on the first
order error term.  This can be set equal to zero by choosing
\begin{equation}
\phi_2=\phi_1 \pm \arccos(-\pi/2\theta_1)
\end{equation}
and, for consistency with equation~\ref{eq:scrof180}, we will use
the minus sign in future. Sequences obeying this equation will be
insensitive to pulse length errors; the rotation and phase angle
can then be adjusted by choosing suitable values for $\theta_1$
and $\phi_1$. As before we will derive values for a $\theta_x$
pulse; pulses with other phase angles can be obtained by
offsetting all the phase angles by the desired amount.

Solving these equations is complex, but the solutions are fairly
straightforward:
\begin{align}
\theta_1=\theta_3&=\arcsinc\left(\frac{2\cos(\theta/2)}{\pi}\right)\\
\theta_2&=\pi\\
\phi_1=\phi_3&=\arccos\left(\frac{-\pi\cos\theta_1}{2\theta_1\sin(\theta/2)}\right)\\
\phi_2&=\phi_1-\arccos(-\pi/2\theta_1)
\end{align}
where $\sinc(x)$ is defined as $\sin(x)/x$.  We refer to this as a
Short Composite ROtation For Undoing Length Over and Under Shoot
or \textsc{scrofulous} sequence.

\begin{figure}
\includegraphics[scale=0.2]{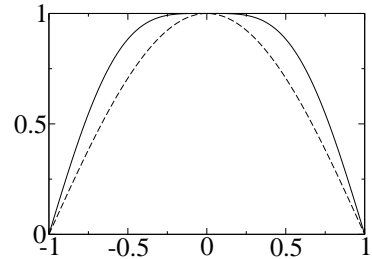}
\caption{Fidelity of simple (dashed line) and \textsc{scrofulous}
composite pulses (solid line) as a function of the fractional
pulse length error $g$ for pulses with a target rotation angle of
$180^\circ$.} \label{fig:scrofg}
\end{figure}

Numerical values of individual pulse rotation and phase angles for
a variety of target angles are given in
Table~\ref{tab:scrofulous}.
\begin{table}
\begin{tabular}{rrrrr}\hline
$\theta$&$\theta_1$&$\phi_1$&$\theta_2$&$\phi_2$\\\hline
$30^\circ$&93.0&78.6&180.0&273.3\\
$45^\circ$&96.7&73.4&180.0&274.9\\
$90^\circ$&115.2&62.0&180.0&280.6\\
$180^\circ$&180.0&60.0&180.0&300.0\\\hline
\end{tabular}
\caption{Pulse rotation and phase angles for a \textsc{scrofulous}
composite pulse with a target rotation of $\theta_x$; note that
$\theta_3=\theta_1$ and $\phi_3=\phi_1$.}\label{tab:scrofulous}
\end{table}
The performance of \textsc{scrofulous} and plain $180^\circ$
pulses are compared in Figure~\ref{fig:scrofg}.

\section{The BB1 family}
Another approach to composite pulse design has been described by
Wimperis \cite{wimperis94}.  While sequences such as
\textsc{corpse} and \textsc{scrofulous} seek a single composite
pulse which performs the desired rotation with reduced sensitivity
to errors, an alternative approach is to combine a naive pulse,
which performs the desired rotation, with a sequence of
error-correcting pulses, which partially compensate for
imperfections.  This approach appears to simplify the design of
composite pulse sequences, and for the case of pulse length errors
produces excellent results.

The form of the error correcting pulse sequence is quite tightly
constrained, as it must have no overall effect in the absence of
errors, but it must also retain sufficient flexibility that it can
act against errors when they do occur.  Here we concentrate on one
particular form suggested by Wimperis \cite{wimperis94}
\begin{equation}
180_{\phi_1}\,360_{\phi_2}\,180_{\phi_1} \label{eq:W1}
\end{equation}
where the values of $\phi_1$ and $\phi_2$ remain to be determined.
When placed in front of a $\theta_x$ pulse Wimperis refers to the
entire sequence as BB1 \cite{wimperis94}, but as we will
generalise his approach we refer to the error correcting sequence
(Equation~\ref{eq:W1}) as W1.

As before we evaluate the quaternion for the composite rotation
(W1 followed by a $\theta_x$ pulse) in the presence of pulse
length errors, and then expand this quaternion as a Maclaurin
series in $g$, the fractional error in the pulse power.  The $y$
and $z$ components in the first order error term are easily
removed by setting $\phi_2=3\phi_1$; the remaining components can
then be eliminated by choosing
\begin{equation}
\phi_1=\pm\arccos\left(-\frac{\theta}{4\pi}\right).
\end{equation}
The positive solution is then identical to that previously
described \cite{wimperis94}.  Examining higher order error terms
shows that this pulse sequence is even better than it first
appears, as these choices also completely remove the second order
error terms.  As discussed below, this effect appears to be a
property of the W1 sequence and its close relations.

It is easy to imagine a range of variations of the BB1 sequence.
Most simply the W1 error correction sequence can be placed
\emph{after} the $\theta_x$ pulse, instead of before it.
Unsurprisingly this has no effect: the solution is the same as
before, and the performance of this reversed sequence is identical
to that of BB1.  More surprisingly the W1 sequence can be place in
the middle of the $\theta_x$ pulse, so that the overall sequence
\begin{equation}
(\theta/2)_x\,\text{W1}\,(\theta/2)_x \label{eq:BB1sym}
\end{equation}
is time symmetric.  Indeed the W1 pulse can be placed at
\emph{any} point within the $\theta_x$ pulse, with almost
identical effects.  The form of the composite quaternion depends
slightly on where the W1 pulse is placed, but the fidelity of the
pulse sequence is unchanged: all error terms below sixth order are
cancelled, with the size of the sixth order term depending on the
value of $\theta$.

\begin{figure}
\includegraphics[scale=0.2]{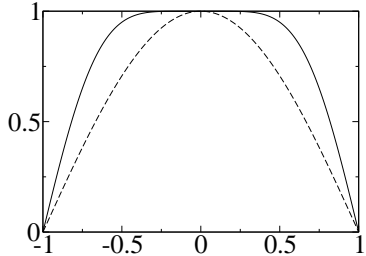}
\caption{Fidelity of simple (dashed line) and BB1 composite pulses
(solid line) as a function of the fractional pulse length error
$g$ for pulses with a target rotation angle of $180^\circ$.}
\label{fig:BB1}
\end{figure}

Numerical values of pulse phase angles for a variety of target
angles are given in Table~\ref{tab:BB1}.
\begin{table}
\begin{tabular}{rrr}\hline
$\theta$&$\phi_1$&$\phi_2$\\\hline
$30^\circ$&92.4&277.2\\
$45^\circ$&93.6&280.8\\
$90^\circ$&97.2&291.5\\
$180^\circ$&104.5&313.4\\\hline
\end{tabular}
\caption{Pulse phase angles for a W1 correction sequence with a
target rotation of $\theta_x$; pulse rotation angles are
$\theta_1=180^\circ$ and $\theta_2=360^\circ$.}\label{tab:BB1}
\end{table}
The performance of the BB1 and plain $180^\circ$ pulses are
compared in Figure~\ref{fig:BB1}.
For all target angles below $180^\circ$ the BB1 composite pulse
outperforms a simple pulse when $|g|<1$.

Another simple variation is to use two or more error correcting
sequences; as before these can be placed at various different
points around or within the $\theta_x$ pulse.  For simplicity we
assume that all the error correcting sequences are identical to
one another, and have the same general form as W1.  In this case
it can be shown that the correction sequences, which we call Wn,
have phase angles given by $\phi_2=3\phi_1$ and
\begin{equation}
\phi_1=\pm\arccos\left(-\frac{\theta}{4n\pi}\right).
\end{equation}
where $n$ is the total number of sequences used.  As before the
fidelity is independent of where the Wn sequences are placed, but
it does depend on the value of $n$.  The second and fourth order
error terms are cancelled in all cases, and the size of the sixth
order error term now depends on both $\theta$ and $n$.  The
smallest sixth order term is achieved when $n=2$, but the term is
not completely removed.  The gain over $n=1$ is fairly small, and
in practice the simpler composite pulses based on the W1 sequence
are likely to be the most effective.

Having varied the position and number of the error correcting
pulse sequences the next logical step is to vary their form.  In
principle any sequence that has no overall effect in the absence
of errors could be used.  In practice we find that many possible
sequences allow the second order error term in the fidelity
expression to be removed, but the simultaneous cancellation of
second and fourth order errors seems to be a special feature of
the Wn family of sequences.

Given the success of this approach to tackling pulse length
errors, it seems obvious to apply the method to tackle
off-resonance effects.  As yet, however, this approach has had no
success.

\section{Simultaneous Errors}
So far we have only considered the case of \emph{either}
off-resonance effects \emph{or} pulse length errors being present.
In reality both problems may well occur simultaneously.  It is
therefore important to consider how such simultaneous errors might
be tackled.  Ideally we would like to design pulse sequences which
can compensate for both problems at the same time; this, however,
is a complicated and as yet unresolved problem, and here we simply
analyse the sensitivity of each of our pulse sequences to the
\emph{other} kind of error.

We proceed as before, calculating composite pulse and simple pulse
quaternions in the presence of errors, and determining the
quaternion fidelity.  This fidelity can then be expanded as a
Maclaurin series in the error, and the lower order terms examined.
Note that this procedure still assumes that only one type of error
is present at a time; in order to detail with the case where both
errors are present \emph{simultaneously} it would be possible to
use a Maclaurin expansion in both errors, but this is unlikely to
lead to much insight.  Instead we will simply plot the fidelity as
function of both errors for some chosen target angle.

\begin{figure*}
\includegraphics[scale=0.9]{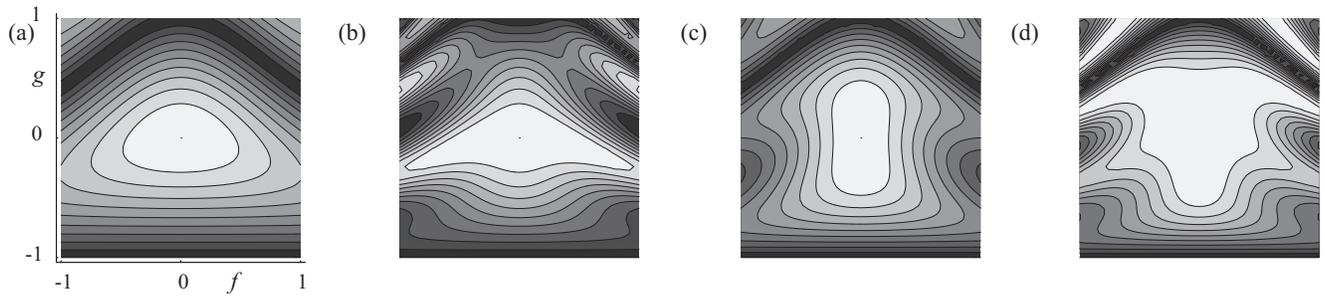}
\caption{Fidelity of (a) plain, (b) \textsc{corpse}, (c)
\textsc{scrofulous} and (d) BB1 $180^\circ$ pulses as a function
of simultaneous off resonance effects, $f$, and fractional pulse
length error, $g$.  Contours are plotted at 5\% intervals.}
\label{fig:2dplots}
\end{figure*}

We begin by considering the response of the \textsc{corpse} pulse
sequence to pulse length errors.  In the absence of off-resonance
effects the behaviour of \textsc{corpse} is trivial to calculate,
as the three pulses are applied along the $+x$, $-x$ and $+x$
axes, so that the behaviour is identical to that of a simple
pulse.  The behaviour of a $180^\circ$ pulse in the presence of
simultaneous errors is shown in Figure~\ref{fig:2dplots}.

The behaviour of the \textsc{scrofulous} pulse sequence is
difficult to calculate for general target rotation angles, due to
the dependence of $\theta_1$ on the $\text{arcsinc}$ function, and
so we concentrate on the case of $180^\circ$ pulses.  For this
case the dependence of the fidelity on off-resonance effects is
given by $\mathcal{F}\approx1-2f^2$, while a simple pulse has a
fidelity $\mathcal{F}\approx1-f^2/2$ (see
Equation~\ref{eq:fidfplain}). In general \textsc{scrofulous} is
considerably more sensitive to off-resonance effects than plain
pulses.

Finally we consider the BB1 family of pulse sequences, taking the
time-symmetrised version of BB1, Equation~\ref{eq:BB1sym}, as our
standard.  In this case we can solve the problem for any target
rotation angle, and up to second order the result is identical to
that of a plain pulse, Equation~\ref{eq:fidfplain}.  Thus, unlike
\textsc{scrofulous}, the BB1 sequence achieves its impressive
tolerance to pulse length errors at little or no cost in
sensitivity to off-resonance effects.  This is confirmed for
simultaneous errors by Figure~\ref{fig:2dplots}.

\section{Conclusions}
Composite pulses show great promise for reducing data errors in
NMR quantum computers. More generally, any implementation of a
quantum computer must be concerned on some level with rotations on
the Bloch sphere, and so composite pulse techniques may find very
broad application in quantum computing. Composite pulses are not,
however, a panacea, and some caution must be exercised in their
use.

The \textsc{corpse} pulse sequence appears to be the best approach
for tackling small off-resonance errors (for large \emph{known}
off-resonance effects the resonance offset tailored, or
\textsc{rotten}, scheme \cite{cummins01b} is preferable).  For
pulse length errors variations on the BB1 scheme of Wimperis
\cite{wimperis94} give the best results; the \textsc{scrofulous}
family of pulses is less effective, but does have the advantage of
being considerably shorter.

\begin{acknowledgments}
HKC thanks NSERC (Canada) and the TMR programme (EU) for their
financial assistance. JAJ thanks the Royal Society of London for
financial support.
%The OCMS is supported by the UK EPSRC, BBSRC, and MRC.
\end{acknowledgments}

\end{document}